# In-Situ TEM Observation of Solid-State Phase Evolution in High-Entropy Alloy Nanoparticles

Nirmal Kumar Katiyar[1], Noorbasha Bhavani Sai[2], Guilherme Gurian Dariani[3], Douglas S. Galvao[4], Pulickel M. Ajayan[5], Pedro Alves da Silva Autreto[6], Sriram Vijayan[7]*, Chandra Sekhar Tiwary[2]*

[1]School of Advanced Engineering, University of Petroleum and Energy Studies (UPES), Bidholi, Dehradun, UK-India-248007

[2]Metallurgical and Materials Engineering, Indian Institute of Technology, Kharagpur, West Bengal, India-721302

[3]Ilum School of Science, Brazilian Centre for Research in Energy and Materials, Campinas, Brazil.

[4]State University of Campinas, Applied Physics Department and Centre for Computing Engineering and Sciences, Campinas, São Paulo, 13083-970, Brazil

[5]Department of Materials Science and Nano Engineering, Rice University, Houston, Texas, USA

[6]Center for Natural and Human Sciences (CCNH), Federal University of ABC (UFABC), 09210-170 Santo André, SP, Brazil

[7]Department of Materials Science & Engineering. Michigan Technological University, Houghton, MI, USA

Email: sriramv@mtu.edu; chandra.tiwary@metal.iitkgp.ac.in;

**Abstract**

The emergence of new materials has gone beyond conventional design principles. Among them are high entropy alloys (HEAs), in which multiple metallic elements coexist within a single unit cell. In this case, each atomic position in a solid material has a different chemical environment, leading to configurational entropy as high as that in the fully random or melted state, which has been frozen to a solid solution. However, the behaviour of such alloy systems at pre-melting temperatures is largely unknown, where atoms gain sufficient mobility and time to rearrange, leading to an enthalpic rather than an entropic interplay. This study demonstrated the phase transformation of HEA nanoclusters (CoCrFeMnNi) under high-resolution transmission electron microscopy (HRTEM) at different temperatures, including room temperature (RT), 550 °C, and 850 °C. The clustered nanoparticles coarsen into larger clusters, accompanied by compositional heterogeneity due to surface and lattice diffusion. This experimental finding of atomic mobility was further corroborated using molecular dynamics, which provides a detailed mechanistic description.

## 1 Introduction

High entropy alloys (HEAs) have attracted significant interest since its discovery in 2004 [1]. The unusual combination of structural and functional properties attainable is attributed to the thermodynamic complexities of a multi-principal elemental system, which can lead to unique local chemical ordering [2-5]. In a typical HEA, when five or more metals with different crystal structures are melted together in a single crucible in equal proportions, the system forms a concentrated, but highly random, solid solution in the liquid state, resulting in high configurational entropy ($\Delta S_{conf}$)[5, 6], which leads to a thermodynamically stable alloy system in the solid state, where the entropic contribution overcomes the enthalpic contribution, minimising the Gibbs free energy of the system and thereby stabilising a disordered single-phase solid solution in these equi-atomic complex concentrated alloys. This is contrary to traditional metallurgical alloying principles, where one primary element usually defines the crystal structure, and adding multiple minor alloying elements produces a dilute alloy that exhibits phase separation, eutectic solidification, ordering, and/or segregation [7]. The equi-atomic CoCrMnFeNi alloy, also popularly known as the Cantor alloy, is a canonical system consisting of a single-phase FCC microstructure that has emerged as a structural material of significant interest due to its attractive combination of strength, ductility and fracture toughness, particularly under cryogenic conditions [8, 9]. The 'four core effects' are used to rationalize the unique combination of properties associated with the emergent behaviour of these alloys; The high entropy effect promotes the FCC single phase, lattice distortion modifies dislocation behaviour, the cocktail effect results in emergent mechanical behaviour due to multi-element interactions that extends beyond simple 'rule of mixtures' assumptions of strengthening and finally, sluggish cooperative diffusion due to severe lattice distortion is associated with the slow kinetics of phase transformations. The latter effect is critical to understanding the microstructural evolution of these alloys during post-process heat treatments. The Cantor alloy was initially believed to exist as a single-phase alloy below its solidus temperature. However, studies have shown that the FCC single phase decomposes when subjected to heat treatments at medium-to-long hold times below 900 ˚C. Otto *et al.* reported the phase stability of a coarse-grained CoCrFeMnNi alloy when exposed to 500 days of isothermal heat treatments between 500-900 ˚C. They observed, at 900 ˚C for 500 days, the FCC phase remained stable, whereas at 700 ˚C and 500 ˚C for 500 days, the single -phase FCC alloy decomposed to a Cr-rich, σ-phase + FCC phase and $L1_0$ (NiMn), B2 (FeCo), BCC (Cr-rich) and FCC phase, respectively[10]. Subsequently, studies have examined the phase stability of annealed and cold-worked Cantor alloy and the microstructure of heat-treated specimens between 500 and 900 °C for durations ranging from 1 h to 720 h [11, 12]. These studies on phase stability showed that the kinetics of phase decomposition at intermediate temperatures in coarse-grained alloys were extremely sluggish due to slow bulk-mediated diffusion. Grain refinement due to severe plastic deformation via high-pressure torsion significantly accelerated the kinetics of decomposition at elevated temperatures due to grain boundary-mediated diffusion[13, 14]. Additionally, the reduced grain size in nanocrystalline thin films of CoCrFeMnNi can significantly delay the onset of secondary phase formation and accelerate its formation due to increased grain boundary density. Thus far, nine different phases have been reported to form, predominantly at grain boundaries due to phase decomposition, which include oxides[10, 12, 15-18], carbides[10, 17], and sulfides, several intermetallic [10, 12, 18] and topologically close-packed phases[17] [12]. However, studies investigating the mechanisms of formation of these phases are limited. Li *et al.*, [19]established that segregation driven grain boundary spinodal decomposition was the precursor mechanism responsible for nucleation of secondary phases in this alloy system. Ming *et al.,* [20]observed co-clustering of Cr and Ni + Mn at grain boundaries (GBs), promoting crack initiation due to reduced cohesion along GBs and resulting in a loss in ductility at high temperatures. These studies suggest the role of interface-mediated nucleation of solute elements during high-temperature post-processing of these alloys. Therefore, approaches that increase interfacial area in these alloy systems will accelerate kinetics of phase decomposition resulting in novel metastable and/or stable phases. Although the phase stability of bulk polycrystalline CoCrFeMnNi alloy is reasonably well understood not much is known about the phase stability of these alloys in its nanoparticulate form.

Recent success in the synthesis of HEA NPs[21, 22] has enabled us to combine the unique nanoscale phenomena observed at these length scales with the 'four core effects' observed in HEAs. This provides a new paradigm for the design of alloyed NPs that cannot be realised via conventionally alloyed or monometallic NPs. HEA NPs have generated significant interest for their excellent catalytic properties, which require high thermal stability[23, 24]. Previously the thermal stability of bimetallic alloy NPs and monometallic NPs have been the subject of several computational and experimental investigations. The latter NPs typically exhibit phenomena including sintering, oxidation, sublimation and/or full melting, whereas the former NPs show, in addition to the phenomena mentioned, segregation-driven phenomena that result in Janus-type NPs[25] and/or core-shell-type NPs[26]. More recently, studies investigating the thermal stability of HEA's were found to broadly fall under three different types of processes; surface-initiated disordering/melting (pre-melting)[27], multi-stage melting[27], enhanced element-specific mobility due to weaker bonding and cohesive energies and size-dependent scaling of melting. However, the time scales of these computational studies are ideally suited for understanding the early stage thermally activated processes and cannot fully capture phenomena such as coarsening, phase separation that occur at experimental time scales of minutes to hours.

Although there have been previous studies investigating the thermal stability of HEA NPs of modifications of alloy systems. Currently, there are no systematic experimental studies on the phase stability of equiatomic CoCrFeMnNi nanoparticles as a function of heat treatment temperature. In this context, several fundamental questions remain unanswered; (1) At what temperature does the metastable solid solution decompose? (2) Which elements segregate, and do they segregate to the surface or the interior of NP? (3) Is the phase separation consistent with bulk equilibrium thermodynamics (CALPHAD), or does the surface energy contribution alter the phase assemblage? (4) What is the kinetic mechanism responsible for the transformation? Surface segregation, GB-analog spinodal, or classical nucleation-and-growth? In situ transmission electron microscopy heating of NPs are an excellent approach to study the phase stability of these complex alloys at high spatial resolution and observe the thermally activated processes responsible for these transformations.

In this report, we investigated the phase stability and the mechanisms governing phase formation in equiatomic FeCoCrMnNi alloy nanoparticles via *in-situ* TEM heating experiments. *In-situ* isothermal heating experiments were performed at 550°C and 850 °C to observe the phase stability of these NPs. Bright Field and High-Resolution TEM images, diffraction patterns and STEM EDXS maps were obtained to identify the crystal structure and estimate the chemical composition of these NPs before and after heating. The experimentally observed phase separation is further rationalized by estimating the HEA's components' mobility at higher temperatures in nanoparticles, as determined by molecular dynamics (MD). to provide mechanistic insights that govern the thermally activated processes in this classical HEA NP system.

## 2 Experimental

### 2.1 Materials and Nanoparticles Preparation

The pure metal buttons of Co, Cr, Mn, Ni, and Fe were purchased from Thermo Scientific Chemicals with a purity of 99.99%. The Co, Cr, Mn, Ni, and Fe were taken in an equi-atomic ratio and melted together in a vacuum arc discharge furnace in a copper crucible. After melting, a 10-g ingot is formed upon solidification. The ingot was flipped and remelted three times to ensure good homogeneity. Then, the ingot was sealed in a quartz tube along with argon gas and annealed at 1000°C for 12 hours. To preserve the homogenised random solid-solution state, the red-hot quartz tube containing HEA ingot is quenched into cold ice water. The annealed ingot was sectioned into smaller pieces and cryomilled by maintaining the milling temperature below -150 $^{o}$C. The Tungsten carbide ball and vial were used for milling. Liquid nitrogen ($LN_2$) was used to maintain the cryogenic milling temperature. However, the liquid nitrogen and milling powder never get mixed due to a separate jacket for $LN_2$[28]. The Ar gas was continuously purged in the milling chamber to prevent moisture condensation and oxidation of nanoparticles. The milling process was performed for 10 hours to form high-entropy alloy nanoparticles.

### 2.2 In-situ TEM imaging and Phase Evolution

The transmission electron microscope was used to study the phase tranformation behaviour of high entropy alloy nanoparticles. The alloyed nano powder was dispersed in ultrapure methanol, and after sonication, the suspension was drop-cast onto the amorphous silicon nitride holey film on the micro electromechanical systems (MEMS)-based heating chip. This MEMS chip is compatible with a single tilt NanoEx -i/v r *in-situ* heating stage. The NPs were heated to 550 $^{o}$C at a ramp rate of 10 $^{o}$C/min, and the sample was held at this temperature for ~20 minutes to observe any changes and then rapidly quenched at 10 C/s . Next, the sample is heated to 850°C at the same rate and held for ~5 minutes. BF-TEM and HR-TEM images were obtained before and after heating at each isothermal hold. The movie was recorded at a frame rate of 1.29 s/frame using the Velox software to analyse the thermally activated processes. Based on a previous temperature calibration study [29] performed on the MEMS-based heating stage similar to the one used in this study, the HEA NPs dispersed across the silicon nitride holey film experience an actual temperature of ~550˚C and 800˚C for a setpoint temperature of 550˚C and 850˚C, respectively. The TEM used for this study is a probe-corrected Titan (Thermo Fisher Scientific Inc.) operated at a maximum accelerating voltage of 200kV. The instrument is attached with a ChemiStem energy dispersive x-ray spectrometer (EDXS) that was used to estimate the chemical composition of the phases observed before and after heating. The STEM EDXS Maps were obtained at a pixel resolution of 512 x 512 pixels for a total acquisition time of ~ 20 mins. The x-ray maps were quantified using the Velox software to obtain quantified compositional maps.

The prepared high-entropy alloy ingot and nanoparticles were characterised using X-ray diffraction (Empyrean Panalytical) with a Cu Kα wavelength of 1.54056 Å for a phase identification and to detect any abrupt changes during milling. The field-emission scanning electron microscope (FESEM) with EDS (JEOL FESEM) was used to study the composition and chemical homogeneity. Furthermore, the Thermo Fisher electron microscope was used in an *in-situ* phase change analysis, and the temperature was controlled using the associated assembly.

### 2.4 Simulation details

The structural and chemical ordering of the synthesized alloyed nanoparticles was investigated using fully atomistic molecular dynamics (MD) simulations. A total of 6 simulated systems were generated, comprising two distinct nanoparticle diameters (approximately 46 Å and 70 Å). For each size, three independent initial atomic distributions were created to ensure statistical rigor and assess size-dependent effects.

The molecular simulations were executed utilizing the LAMMPS (Large-scale Atomic/Molecular Massively Parallel Simulator) package[30]. The system was modelled within the canonical (NVT) ensemble [NVT][31], maintaining temperature via a Nose-Hoover thermostat, executed using the KOKKOS [32] accelerator package for performance optimization. Interatomic interactions were represented by the Modified Embedded Atom Method (MEAM) force field [33], which has been validated for similar metallic systems.

Periodic boundary conditions (PBC) were implemented. The dimensions of the simulation cell were selected to ensure a minimum separation distance of 15Å between the nanoparticle and its periodic images. Nanoparticles were initially generated with a random elemental distribution, followed by a structural relaxation prior to the main thermal cycle. A detailed thermal protocol was implemented to simulate the formation and internal reorganization processes. The thermal protocol consisted of a stepwise ramp: the temperature was raised from an initial 300 K to a maximum of 2500 K in 100 K increments, and subsequently cooled to a final 300 K at a rate of 40 K per step. Each temperature interval was achieved via a 100 ps ramp, immediately followed by a 100 ps thermalization/equilibration run at the final temperature of that step, ensuring adequate atomic rearrangement and the formation of stable configurations.

## 3 Results and discussion

To investigate the phase transformation behaviour, a nanocrystalline Cantor alloy was prepared by casting, followed by cryomilling [22, 34]. The microstructure of the high-entropy alloy after casting and subsequently cryomilling to alloyed nanoparticles was found to consist of the face-centred cubic (FCC) phase; no phase change is evident, as the XRD pattern of the HEA ingot when converted to nano powder as shown in **Supplementary Figure S1**. Typically, the HEA's nanoparticles have a tendency to oxidize, however, the ball milling at cryogenic temperature suppresses the oxidation of nanoparticles [22, 35]. The FCC phase in the Cantor alloy (FeCoCrMnNi) is a complete solid solution, with all five elements randomly distributed in the solid state. A key question that we address here is how this fully random solid solution of five elements behaves at the nanoscale upon exposure to elevated temperatures? Notably, the melting points ($T_m$) of the individual bulk metals, in increasing order, are Mn (1246 °C) < Ni (1455 °C) < Co (1495 °C) < Fe (1583 °C) < Cr (1907 °C). The element with the lowest melting point is manganese (Mn). The weighted average melting point of the bulk high entropy alloy is approximately 1526 °C, and the experimentally estimated bulk melting point is 1334 ºC[36]. To investigate phase stability at elevated temperatures, the FeCoCrMnNi nanoparticles were exposed to isothermal holds at setpoint temperatures of 550 °C and 850 °C, which are consistent with temperatures at which phase decomposition has been observed in previous reports on polycrystalline Cantor alloys. The NanoEx-i/v single-tilt heating holder, compatible with a MEMS-based microheater stage and with excellent thermal stability and temperature homogeneity, was used for in situ isothermal heat treatments inside the TEM. Before heating, the FeCoCrMnNi nanoparticle agglomerate ~250 nm in width exhibited a complete solid solution, as confirmed by quantitative elemental maps, as shown in **Figure 1(a)**. Upon heating the sample to 550°C and isothermally holding for ~10 minutes, Fe and Mn were observed to have elevated surface concentrations along the periphery of the agglomerate, as evidenced by elemental maps in **Figure 1(b)**. The Fe and Mn exhibit elevated atomic percentages at the cluster edges, likely due to the early surface-driven segregation due to pronounced mobility of these elements at ~ $0.5T_m$ ($T_m$ = Melting point) of the nanocrystalline alloy. The NP agglomerate consists of an ensemble of nanocrystalline grains of 5-20 nm in size, which significantly increases the surface area per unit volume, and most atoms are located close to the surface, which acts as a diffusion highway. Bulk diffusion in FCC lattice of the Cantor alloy of these NPs is typically expected to be sluggish. However, grain-boundary and surface diffusion are several orders of magnitude faster than bulk diffusion. Therefore, these boundaries will enable modest segregation of Fe and Mn at 550 ºC without much bulk diffusion. These diffusional processes proceeded along the interfaces of the nanocrystalline grains within the agglomerate, initiating the coalescence of these grains along grain boundaries and triple junctions. Rationalizing the *in-situ* TEM observations from the current study, based on previous bulk tracer diffusion experiments of constituent elements in Cantor alloy[37], which have shown that Mn is the fastest and Ni is the slowest diffusing species, while Cr and Fe have similar diffusion rates and activation energies, which suggests that Fe is not dramatically slower than Cr. Previously, ab-initio calculations by Ferrari *et al.* on CoCrFeMnNi in vacuum have revealed that Mn has a strong tendency to segregate in the outermost layer of the free surface in vacuum, whereas Fe has a weaker but non-negligible tendency to segregate compared to Ni and Co[38] Additionally, they also noted Cr has a stronger tendency to segregate, but is usually restricted to the near surface region. These previous investigations give merit to our *in-situ* TEM observations at 550 ºC, which indicate strong surface-driven segregation of Mn and Fe, which decorate the periphery of the NP agglomerate. Subsequent heating to 850 °C and isothermally holding for ~5 mins revealed the phase transformation, in which the smallest crystallites undergo coalescence and grow into larger particles. The coalescence leads to pathways for accelerated phase separation in the solid solution. Compositional heterogeneity is evident in **Figure 1(c)**. Two dominant regions are evident from the Z-contrast of the HAADF image, as shown in the combined quantified elemental maps in **Figure 1(c)**. The chemical maps revealed a Cr-Mn rich and Ni-Co rich regions. The Cr-Mn appears to form around the Ni-Co rich region, indicating that the Cr-Mn rich phase forms along the periphery of the sintered grains within the agglomerate.

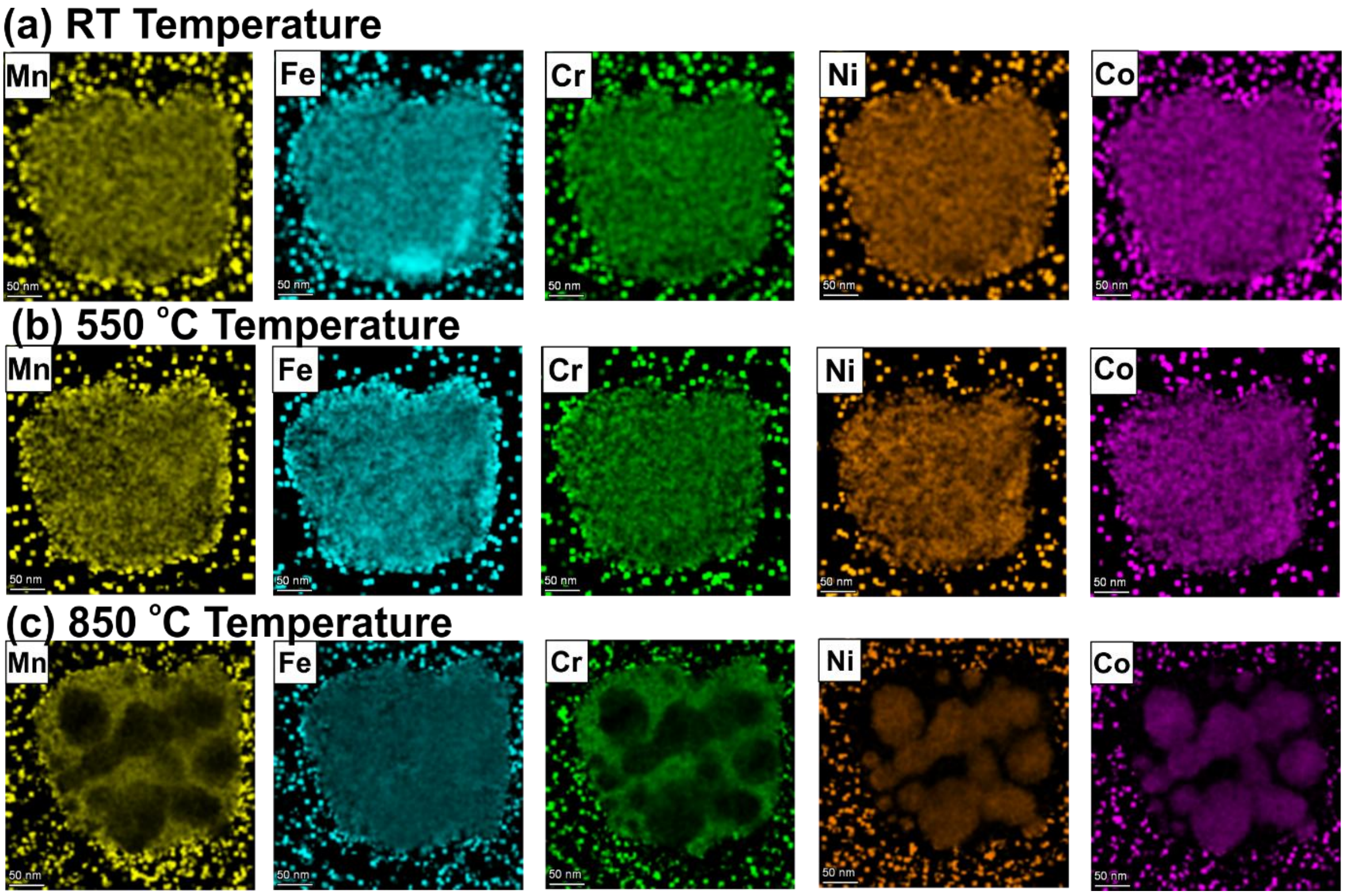

**Figure 1**: *Phase transformation of high entropy alloy FeCoCrMnNi at Room temperature (RT), at 550 °C: and at 850 °C, HAADF colour mapping image shows the combined elements distribution in the nanoparticle of individual atomic distribution of each element (Fe, Co, Cr, Mn, Ni).*

Further, to confirm the phase transformation and compositional heterogeneity across the nanocluster, line-profile composition estimation was performed at room temperature, 550 °C, and 850 °C. The elemental line profile across the homogeneous region is consistent with a regular disordered solid solution of FeCoCrMnNi at room temperature [**Figure 2(a)]** ; however, the surface is non-uniform due to minor variations in the thickness of the particle, leading to differences in intensity at different locations, as seen from the dotted line highlighting the HAADF intensity line profile in **Figure 2(a)**. The isothermal heating of the agglomerate at 550 °C is shown in **Figure 2(b)**, and it is clear that elevated concentrations of Mn and Fe occur at the edge of the nanocluster about ~15 atomic percentage, which is also proven in elemental maps in **Figure 1(b)**. There is no deficiency of Fe and Cr was occur centre of the cluster, which might be due to crystallite surface atoms diffuse through grain boundaries only at this temperature. The *in-situ* movie has been recorded for the phase transformation at 550 °C, as shown in **supplementary movie 1**, and no change was observed. However, at 850 °C, there is clear evidence of phase transformation in the solid state, as seen in **supplementary movie 2**. To confirm the different phase decomposition at 850 ºC, a line profile-concentration was performed across two phase regions, as shown in **Figure 2(c)**. Line scans were obtained from a region spanning both regions, revealing that the Fe content remains constant across them. The Mn and Cr concentrations were depleted in the centre of the one-phase region and elevated in the surrounding region [**Figure 2(c)**]. Additionally, areas of interest 1 and 2 (indicated by the red boxes) were analysed to estimate the composition of the two chemically distinct regions shown in **Figure 2(c).** In the Ni-Co rich region (bright contrast in HAADF), we see most of the Cr and Mn depleted, indicating the formation of a phase tending toward $(Co,Ni)_{0.8}Fe_{0.2}$. Around this region, the Cr and Mn elemental composition shows an enrichment, indicating the formation of a phase tending to $(CrMn)_{0.8}Fe_{0.2}$-rich. Based on the STEM-EDXS quantification maps the nominal $(Co,Ni)_4Fe$ and the $(Cr,Mn)_4Fe$ regions contain a residual $< 0.1$ atomic fraction of the complementary elements Cr +Mn = 0.08 and Ni+Co =0.03 in the Co/Ni rich and Cr/Mn phases respectively. During EDXS acquisition of the map, the STEM probe samples through more than one phase in many pixels, leading to cross-spectral contamination. This, combined with the uncertainties in quantifying low concentrations of 3d transition elements due to overlapping spectral lines, could lead to an overestimation of the true measure of elemental solubility. The in situ HAADF STEM movie captured the formation of the chemically heterogeneous regions coalescing in real-time, as shown in supplementary **movie 2.** The FCC single phase decomposed into a two-phase region, clearly indicating that the enthalpic contribution outweighed the entropic contribution at 850 °C. The nominal $(Cr,Mn)_4Fe$ phase is closer to the previously reported Cr-rich σ phase a while surrounding the nominal $(Co,Ni)_4Fe$ phase.

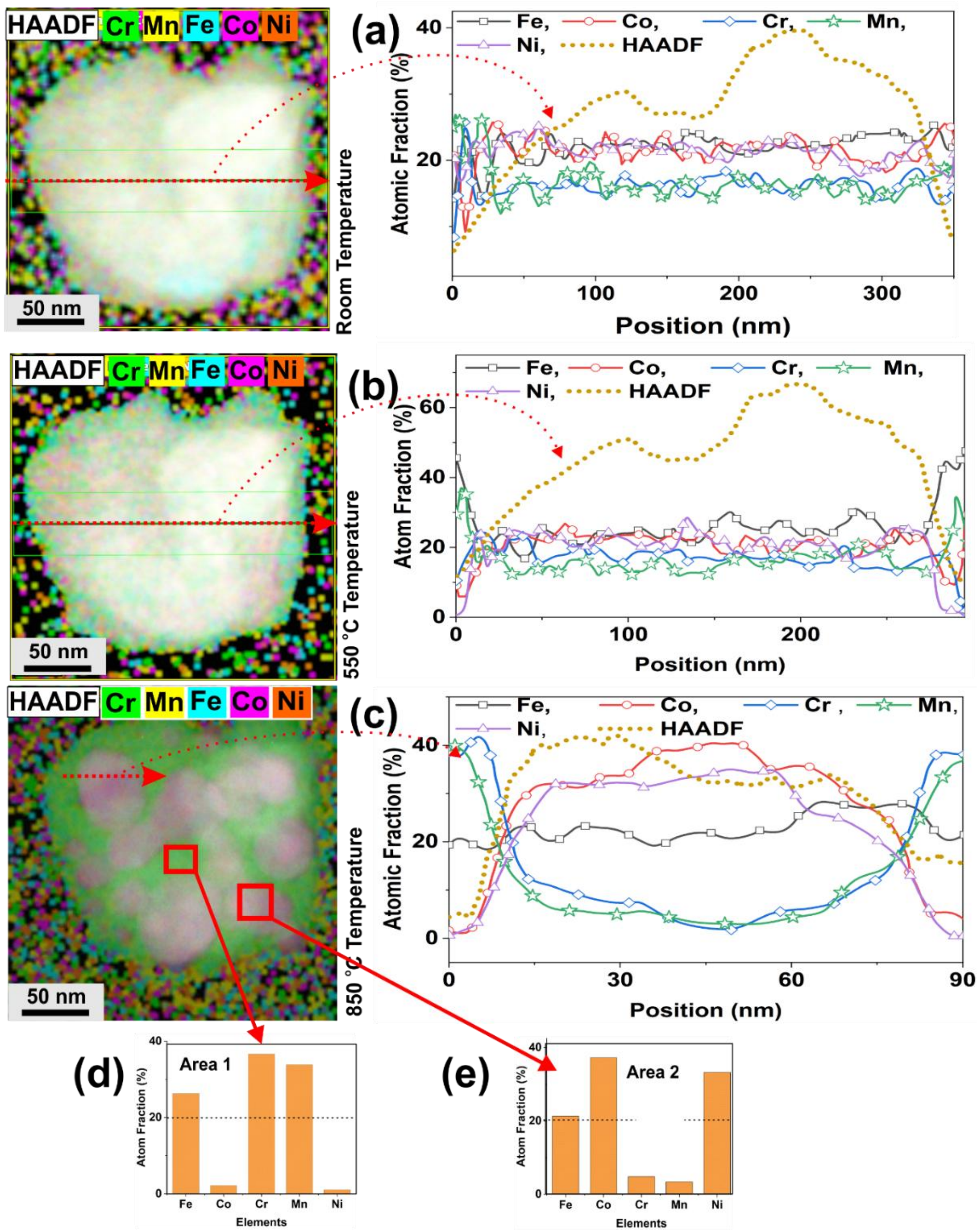

**Figure 2**: *Elemental mapping of the FeCoCrMnNi high entropy alloy nanoparticles at different temperatures and their composition estimation by line scan,* **(a)** *at Room temperature,* ***(b)*** *at 550* $^oC$ *and* ***(c)*** *at 850* $^oC$*,* ***(d)*** *composition of all five elements in area 1* ***(e)*** *composition of five elements in area 2.*

Real-time monitoring of *in-situ* phase transformation is shown in **Figure 3**. A typical behaviour of small crystallites growing into larger particles with heterogeneous compositions is shown schematically in **Figure 3** (five-step schematics). A polycrystalline aggregate having different-sized crystallites, the small crystallites' surface atoms gain mobility at 850 ºC and then slightly larger ones. The thermally activated diffusion between adjacent crystallites is dominant in several smaller crystallites, driving mass transport toward larger crystallite, which then merge and form larger particles[39]. The particles begin to fuse through neck formation and subsequent coalescence[40]. During this process, enhanced atomic mobility at elevated temperatures drives mass transport and disintegrates homogenized particles into two phases due to enthalpic contributions dominating over entropic effects, as the system remains well below the melting point. A subsequent HAADF TEM micrograph of the phase transformation sequence is shown in **Figure 3**.  The small crystallites subsequently merge, followed by larger particles merging through neck formation and coalescence, which is marked by the red dotted lines [**Figure 3**]. A similar phase transformation is observed in a subsequent HAADF image of another cluster, shown in **Figure S3**.

Further, after heating to 850 °C and cooling to room temperature, bright field TEM images along with corresponding selected area diffraction patterns and high-resolution transmission electron microscopy (HRTEM) shown in **Figure 4** were acquired from a large area to compare the changes in the crystal structure. The **Figure 4 (c-e)** shows high-resolution lattice fringe images of different regions from image **4(a).**

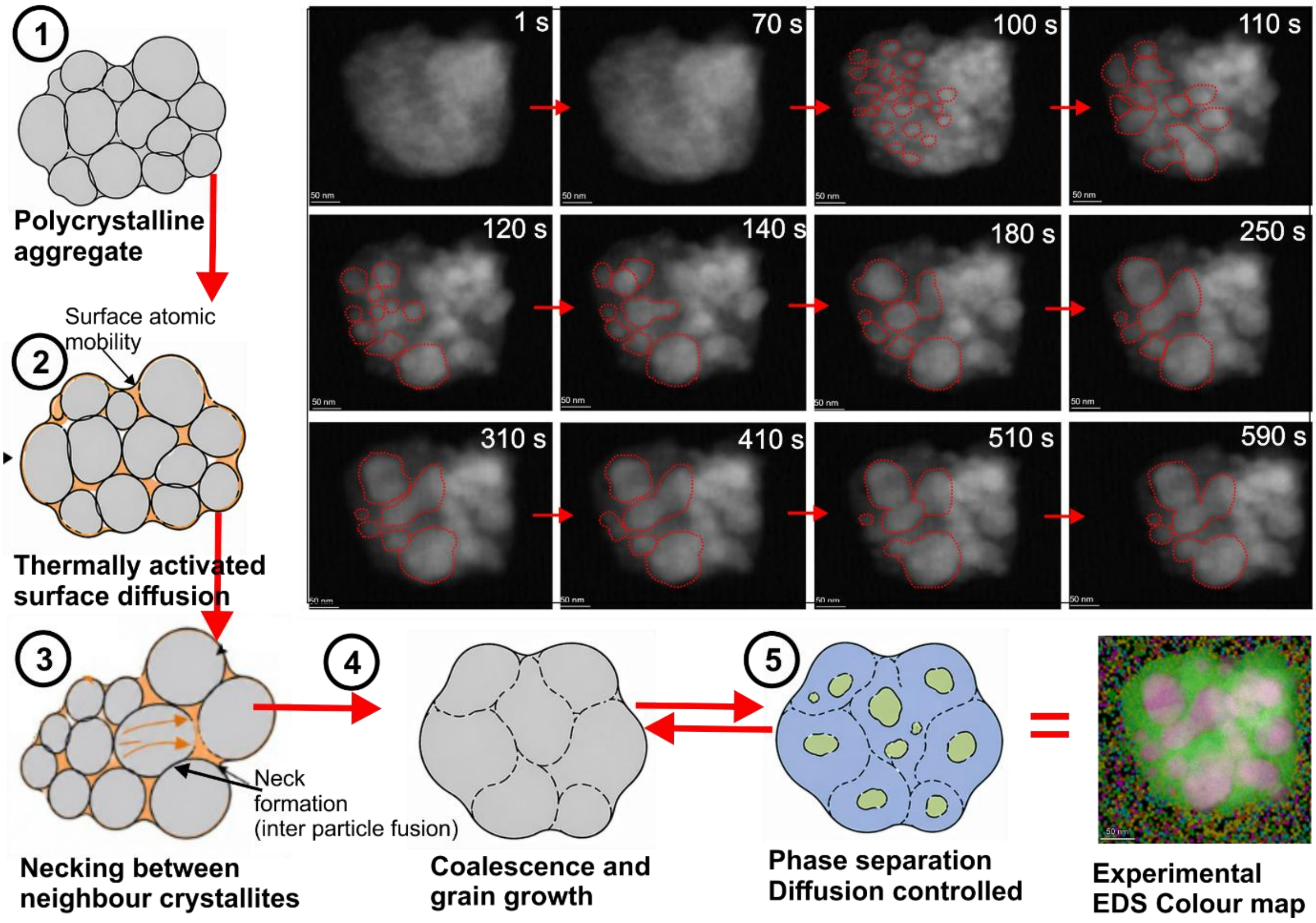

**Figure 3:** ***Phase change at elevated temperature:*** *Schematic of nanocluster thermal surface diffusion, necking, grain growth and phase separation (marked steps 1-5). Real-time in-situ monitoring of nanoparticles, subsequent HAADF image, and phase transformation snapshots at 850 $^oC$.*

The ring diffraction pattern from the region of the micrograph [**Figure 4(a)**] shows that the innermost ring belongs to the mixed metal oxides, which might be from surface metal oxides formed during drop casting and/or formed by oxidation of the metal nanoparticle surface due to minute amounts of residual air in the column at 850 ˚C. However, other rings corresponded to an FCC phase. The SADP also revealed that other closely spaced rings adjacent to the FCC phases, which are marked with a dotted arrow. The SADP was used to confirm the presence of (200), (220) and (311) planes consistent with an FCC phase. These SADPs confirm the presence of two crystallographically similar but chemically distinct product phases, as confirmed by the STEM EDXS chemical mapping in **Figure (1 and 2)**.

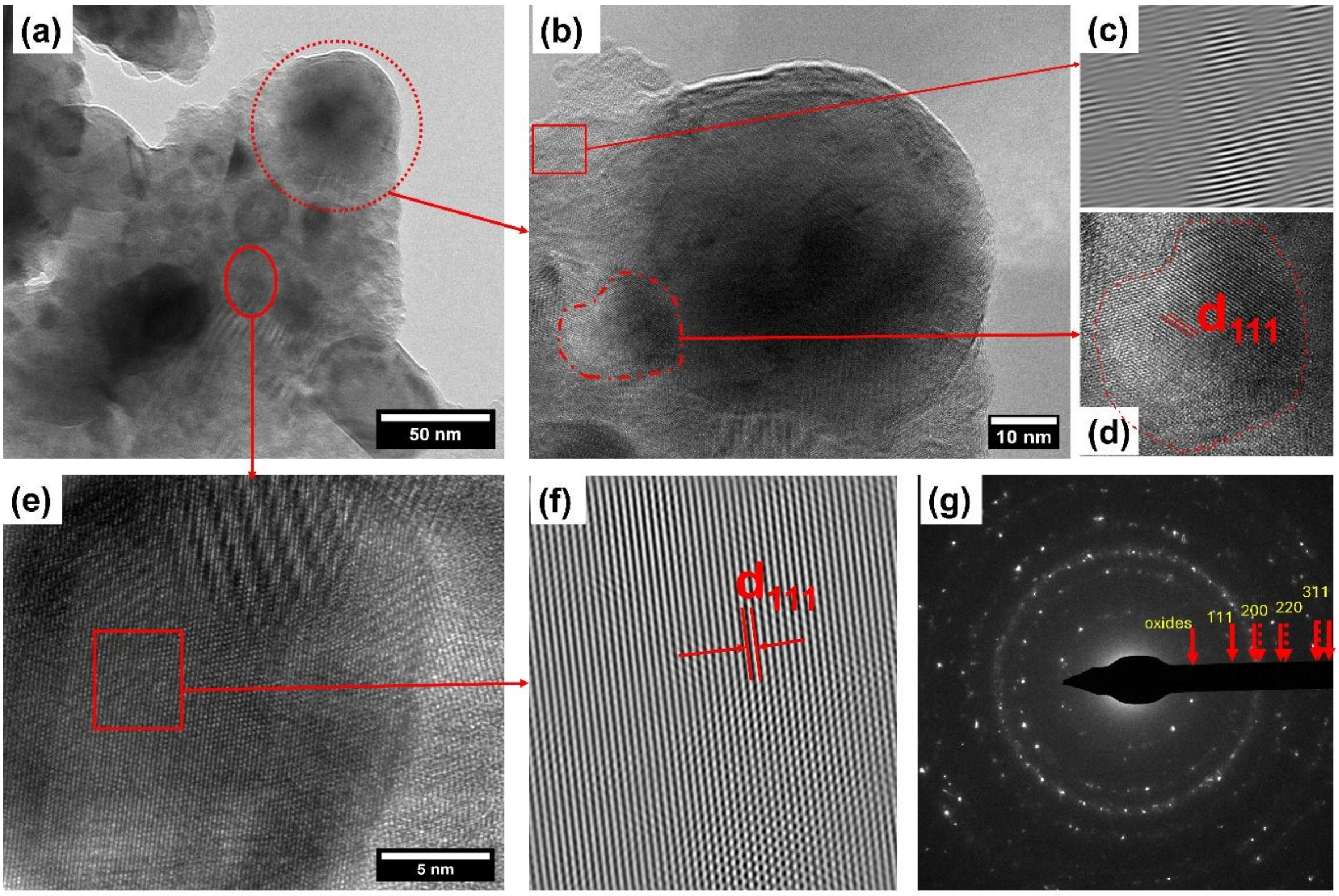

**Figure 4**: High resolution transmission electron micrograph of HEA nanoparticles after heating 850 ºC, ***(a)*** *bright field TEM micrograph of FeCoCrMnNi alloy nanocrystal,* ***(b)*** *magnified view of nanocrystal* ***(c)*** *HRTEM lattice fringes of a crystal from the peripheral of nanoparticles,* ***(d)*** *Lattice fringe from the inner of the particles,* ***(e)*** *lattice fringes of single nanocrystal,* ***(f)*** *FFT filtered lattice fringes,* ***(g)*** *ring diffraction pattern of nanocluster image a.*

It is well established that different temperatures can induce phase transitions that affect the stability of the alloy system. The temperature-dependent phase evolution of the CoCrFeMnNi high-entropy alloy (HEA) deduced from the thermodynamic calculations at different temperatures is presented in **Figure 5(a)**. The results indicate the occurrence of two-phase transitions at 490.9 °C and 667.5 °C. At the beginning of the investigated temperature range (400 °C), the HEA exhibits the coexistence of three phases: an FCC phase (α-FCC), a Co- and Fe-rich BCC phase (β-BCC), and a Cr- and Mn-rich BCC phase (γ-BCC) [**Figure 5(a)**]. Among these, the α-FCC phase is the dominant phase within this temperature range. This three-phase coexistence persists from 400 °C up to 490.9 °C, as shown in **Figure 5(a)**. Beyond the 490.9 °C, a phase transition occurs in which the γ-BCC phase transforms completely into the β-BCC phase, resulting in a two-phase system consisting of α-FCC and β-BCC. This two-phase region remains stable between 490.9 °C and 667.5 °C. Within this temperature interval, the fraction of the α-FCC phase gradually increases, while the fraction of the β-BCC phase decreases. Above 667.5 °C, the alloy transitions into a single-phase structure, where the system stabilizes as an FCC phase. In this region, the HEA exhibits a homogeneous FCC structure, with all constituent elements present in approximately equal proportions. The compositional variations of the phases with temperature were also analyzed and illustrated in **Figure 5(a)**. The α-FCC phase is primarily enriched with Ni and Mn, along with smaller fractions of Co, Fe, and Cr. The proportions of these elements change with temperature up to 667.5 °C, after which they approach nearly equal distribution within the single FCC phase. In contrast, the β-BCC phase initially contains significant fractions of Fe and Co; however, their mole fractions decrease with increasing temperature and approach negligible levels beyond 490.9 °C. The Cr content in the β-BCC phase increases up to 490.9 °C and subsequently shows a declining trend. The γ-BCC phase is initially enriched in Cr and Mn, but the mole fractions of these elements decrease with increasing temperature up to 490.9 °C. At this temperature, the γ-BCC phase transforms completely into the β-BCC phase. Beyond 667.5 °C, the HEA stabilizes as a single-phase FCC structure. However, experimental evidence from **Figures 1 and 2** demonstrates the two-phase decomposition of nanoparticles at 850 °C. This is a phenomenon similar to that observed in experimental and thermodynamic calculations, but discrepancies in the temperature range might be due to size differences and cold-worked nanoclusters.

Furthermore, similar phenomena have been corroborated by molecular dynamics (MD) simulations. The molecular dynamics (MD) simulation investigated the behaviour of synthesised alloy nanoparticles under thermal stress, elucidating the underlying mechanisms of elemental reorganisation. **Figure 5(b-c)**, resulting from the simulation, details the atomic migration of the constituent elements (Fe, Cr, Ni, Co, and Mn) relative to their initial, uniformly distributed atomic states. Study demonstrates a thermally induced elemental reorganization throughout the volume of the nanoparticle, with Iron (Fe) and Manganese (Mn) exhibiting the most significant reorganization, as shown in the **Figure 5(c)**. The applied thermal cycle, simulating synthesis or post-processing conditions, acted as the driving force for this atomic movement. These computational analyses corroborate and provide a mechanistic explanation for experimentally observed phenomena, such as the formation of elemental compositional heterogeneity in the nanoparticles, thereby facilitating the identification of the specific thermodynamic and kinetic driving forces responsible for the observed elemental phase separation. This observed high mobility and inward migration of Fe, coupled with the outward migration of Mn, is consistent with and thermodynamically driven by the highly negative $\Delta E_{_pot}$ for Iron (Fe). This negative potential energy change indicates that Fe-rich core formation is the primary driver for global energy minimization (**Figure S4** in the Supplementary Material).

Further visual confirmation of the final atomic arrangement is provided in **Figure 5(c)**, which shows a snapshot of all atoms within the nanoparticle at the end of the MD simulation. This figure accentuates the differential migration patterns of manganese and iron atoms. Specifically, the simulation reveals that iron (Fe) atoms exhibited a tendency to migrate preferentially towards the central core of the nanoparticle, whereas manganese (Mn) atoms showed a propensity to move towards and concentrate at the external surface, which was also observed by experimental analysis. Through simulation, the nanoparticles' composition heterogeneity is quantified through the atomic concentration analyses presented in **Figure 5(d)**. It provides a detailed profile of the elemental concentration as a function of the radial position, from the core to the surface, for all 6 independent nanoparticles simulated. The figure includes a solid coloured line, representing the average concentration profile, for both the initial (pre-thermal cycle) and final (post-thermal cycle) states of the nanoparticle. The data confirm the qualitative observations: manganese (Mn) atoms preferentially migrate to larger radii (the surface), leading to a manganese-rich outer shell, as evidenced by experimental observations. In addition, the iron (Fe) atoms concentrate in the lower radii (the core), resulting in an iron-enriched core region. This elemental partitioning is a key characteristic defining the final nanostructure and influencing its overall properties. Furthermore, the analysis of the conditional neighbourhood probability (**Figure S5** in the Supplementary Material) provided insight into the local bonding environment, revealing the tendency for Mn and Cr clustering in the surface layer, corroborating the observed elemental segregation. It is worth noting that the experimental observations also support the formation of MnCr-rich phase, confirmed in Figure 2, area-1 composition $(Cr_{0.37}Mn_{0.35}Fe_{0.24})Co_{0.02}Ni_{0.01}$. Therefore, the other region depleted with Mn and Cr and from a different phase, which is CoNi rich.

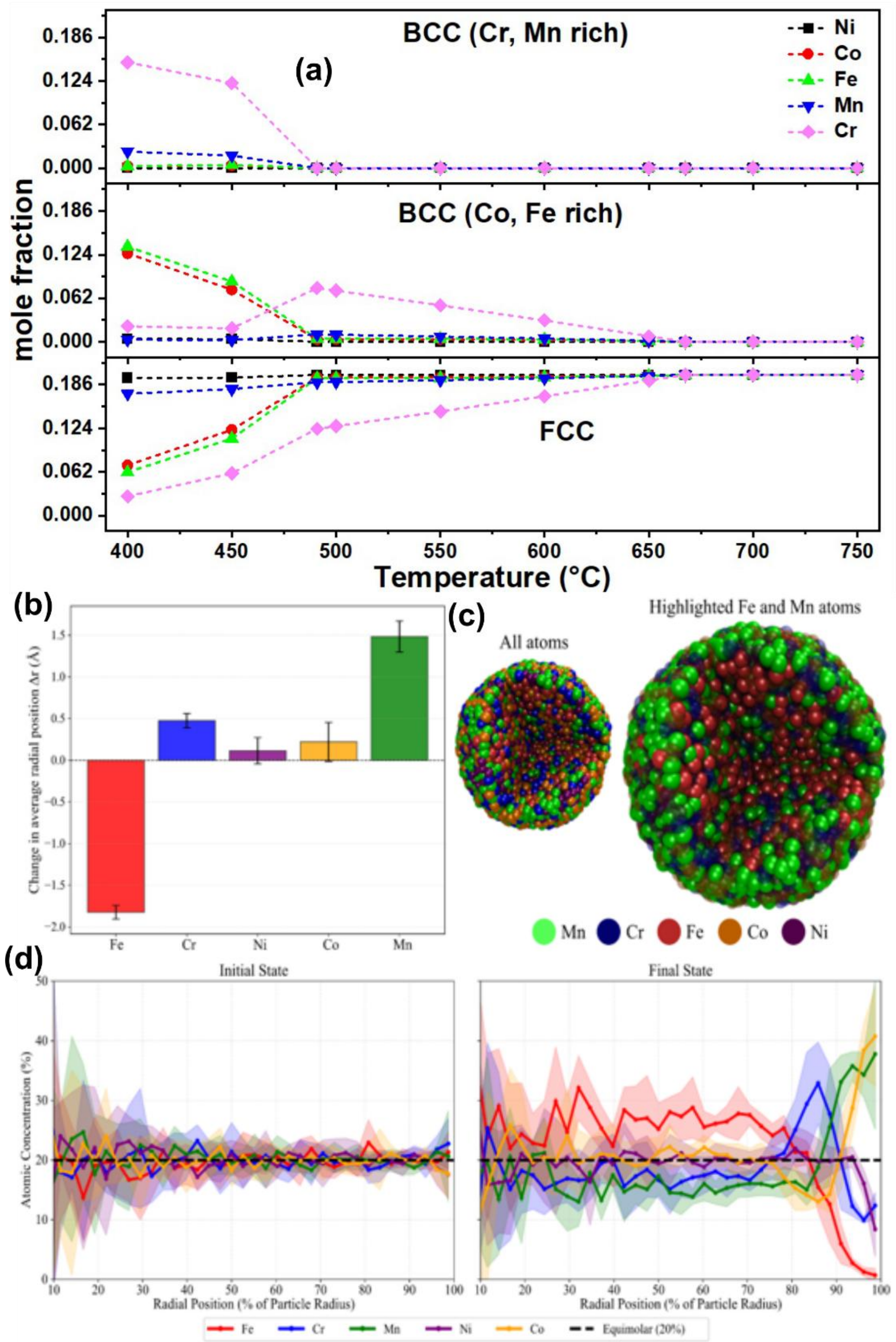

**Figure 5:** *Molecular Dynamics (MD) Simulation Results of the Alloy Nanoparticle.* ***(a)*** *Thermal atomic migration of the constituent elements (Fe, Cr, Ni, Co, Mn) relative to their initial, uniformly distributed atomic states, detailing the thermally induced elemental reorganization.* ***(b)*** *Final atomic configuration of the nanoparticle, accentuating the differential migration patterns of iron (Fe) atoms towards the central core and manganese (Mn) atoms towards the external surface.* ***(c)*** *Quantified atomic concentration profiles as a function of radial position (from core to surface) for all 6 simulated nanoparticles, comparing the initial (pre-thermal cycle) and final (post-thermal cycle) states, confirming the segregation of Mn to*

*larger radii (shell) and Fe to lower radii (core).*

**Conclusion**

The study presents the *in-situ* phase transformation of nanocrystalline CoCrFeMnNi high-entropy alloys using electron microscopy and corroborates the experimental phenomena with the molecular dynamic's simulation. The nanocrystalline cantor alloy undergoes a phase change at 850 °C, inducing compositional heterogeneities within nanoclusters. The diffusion-controlled mass transport was observed during necking and coalescence, which leads to two phases, one is $(Cr_{0.37}Mn_{0.35}Fe_{0.24})Co_{0.02}Ni_{0.01}$, and another is $(Co_{0.36}Ni_{0.34}Fe_{0.21})Cr_{0.05}Mn_{0.03}$. This phenomenon is also supported by thermodynamic calculations at low temperatures, well below the melting point. However, there are discrepancies in the temperature between the thermodynamic calculation and the experimental results. Above 500 ºC, thermodynamics indicates a complete FCC single phase, but experimentally, 550 ºC shows no change in phase transformation. This variation might be due to different-sized crystallites and the strain induced by cold work during ball milling. The Mn-Cr clusterisation was also found by the molecular dynamics calculation at 850 ºC.

**Acknowledgements**

The author would like to thank Anusandhan National Research Foundation (ANRF), India.

**Conflicts of Interest**

The authors declare no conflict of interest

**Data Availability Statement**

The data that support the findings of this study are available from the corresponding author upon reasonable request

**CRediT authorship contribution statement:**

Dr Tiwary has led this work and also conducted materials synthesis and characterisations. Dr. Pedro led in theoretical calculation and data interpretation.

**Associated Content**

i) **Movie 1:** Isothermal heating at 550 ºC
ii) **Movie 2:** Phase transformation 850 ºC

# Supporting Information

1. **X-ray diffraction**: The X-ray diffraction pattern of the high entropy alloy ingot prepared by arc-melting, and then nanoparticles were synthesised by mechanical milling. The X-ray diffraction pattern reveals the face-centred cubic (FCC) crystal structure, as shown in **Figure S1**. The prepared nanoparticles don't show any change in crystal structure. However, the broadening, as shown in **Figure S1** inset for (111) peak, is due to nanoparticle formation.

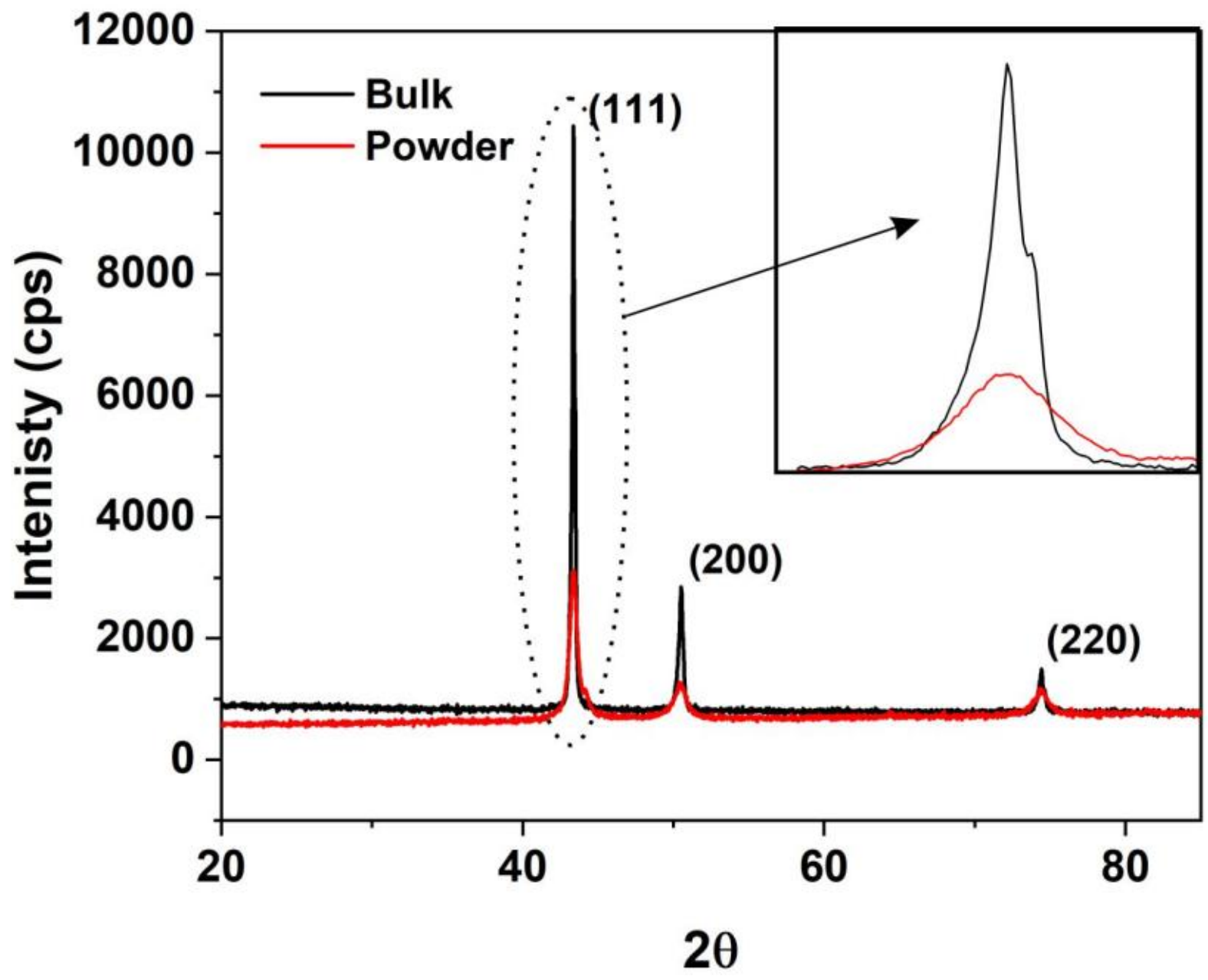

**Figure S1: X-ray diffraction pattern**: XRD pattern of bulk and cryomilled HEA powder, inset shows the broadening in the peak of (111).

2. **Compositional analysis:** Scanning electron microscopy (SEM) were used to assess the composition of the prepared high entropy alloy ingots using associated energy dispersive spectroscopy (EDS). The area scan and point scan, which are marked in **Figure S2**. Both the area and points scan in the materials show similar compositions.

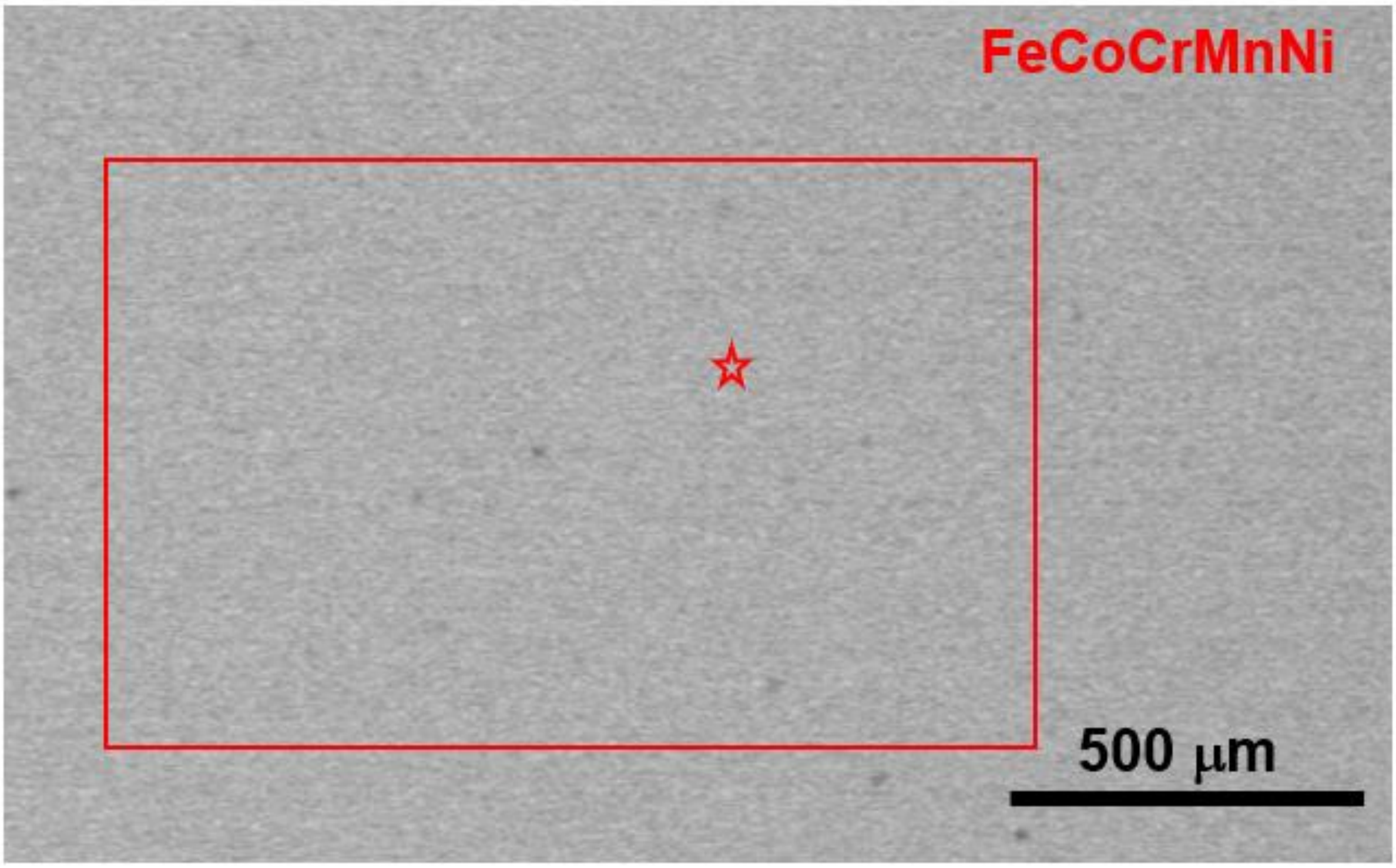

| Scan | Elements | Atoms % |
|---|---|---|
| Area scan (Red line rectangle) | Fe | 19.20 |
| | Co | 20.21 |
| | Cr | 19.33 |
| | Mn | 20.19 |
| | Ni | 21.07 |
| Point Scan (red star marked) | Fe | 19.01 |
| | Co | 19.93 |
| | Cr | 20.32 |
| | Mn | 19.57 |
| | Ni | 21.17 |

**Figure S2: SEM Image of FeCoCrMnNi high entropy alloy**: Energy dispersive Spectroscopy composition analysis, area scan and point scan. The table shows the elemental composition of the red line marked point and area scan.

3. **In-situ phase change in nanocluster:** Successive heating snapshots have been shown in Figure S4. The tiny crystals fused and slowly grew large.

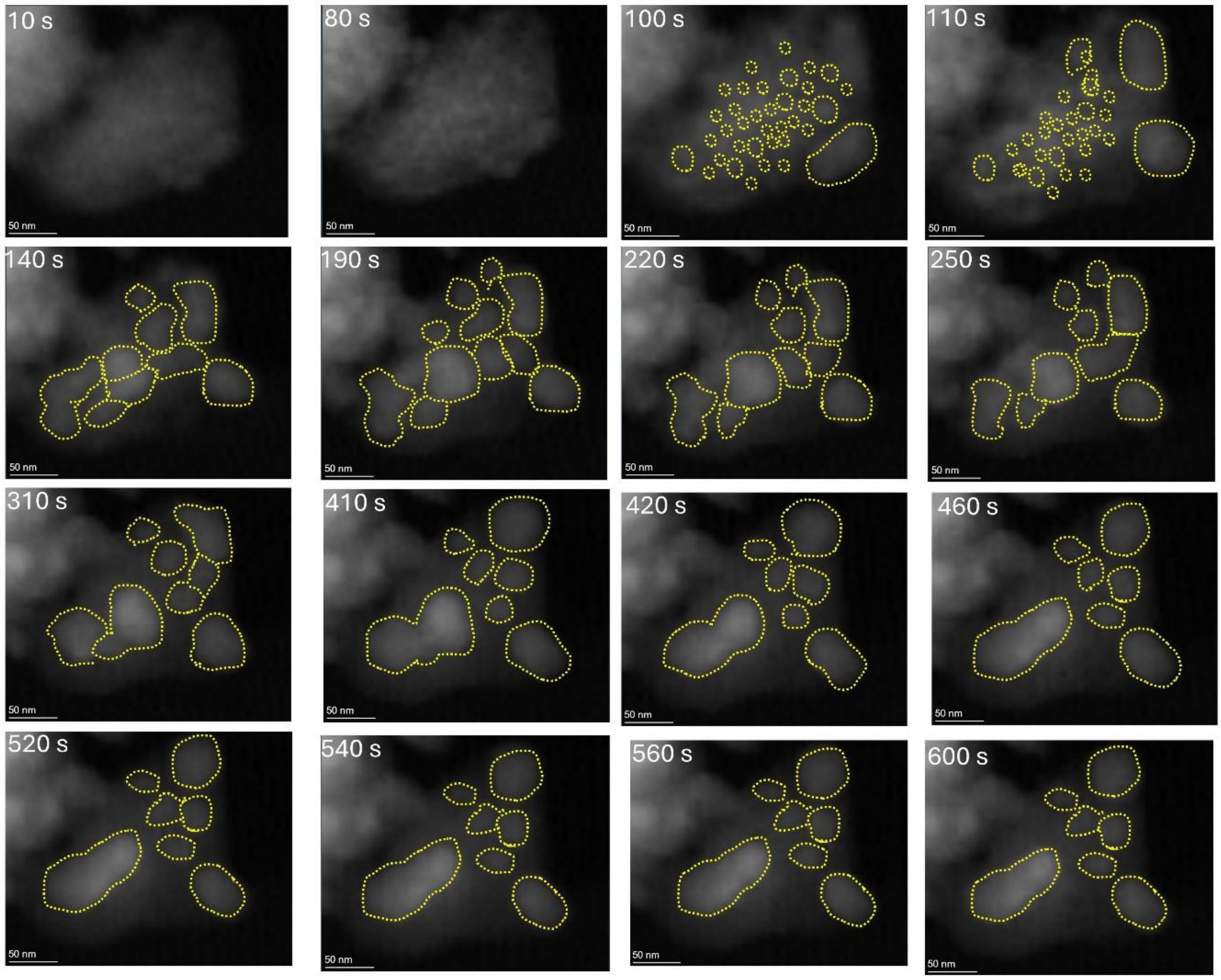

**Figure S3**: TEM bright field micrograph of CoCrFeMnNi alloy nanocluster: Another nanocluster subsequent melting snapshot.

## 4. Thermodynamic Stability Analysis

To quantify the driving forces behind the observed elemental segregation, the change in average potential energy per atom ($\Delta E_{pot}$) was calculated. This metric represents the energy difference between the final segregated configuration and the initial random solid solution. As shown in Figure S4, Iron (Fe) exhibits the most significant negative $\Delta E_{pot}$ value, indicating a substantial gain in stability upon migrating to the core. This identifies the formation of the Fe-rich core as the primary thermodynamic driver for the global energy minimization. Conversely, elements that segregate to the surface exhibit less stable (or positive) energy changes relative to the bulk, reflecting the energetic cost associated with the lower coordination environment of the shell.

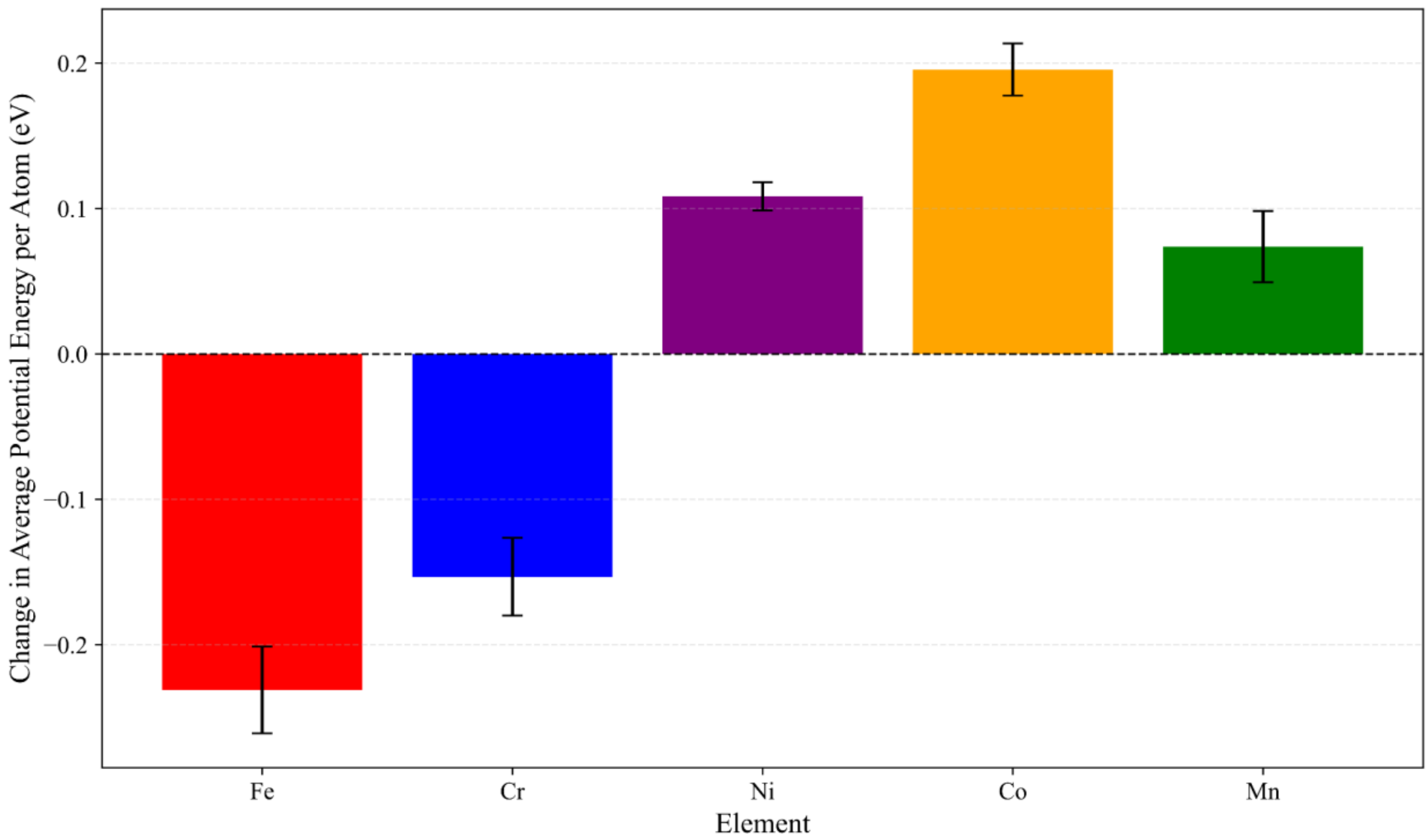

**Figure S4:** Change in the average potential energy per atom ($\Delta E_{pot}$) for each constituent element after the thermal cycle. The significant stabilization (negative energy change) of Iron (Fe) confirms it as the structural anchor of the nanoparticle core.

## 5. Chemical Short-Range Order (CSRO)

The local atomic environment was characterised using conditional neighbour probability analysis across three volumetric regions: Core (0-50% radius), Middle (50-90%), and Shell (90-100%). **Figure S6** presents the probability of finding a specific neighbour type given a central atom. The data reveal a strong affinity for Fe bonds within the core region, exceeding the random distribution probability (dashed line). In the shell region, a distinct clustering tendency is observed for Mn, indicating that the surface layer forms chemically ordered Mn-rich domains rather than a homogeneous mixture.

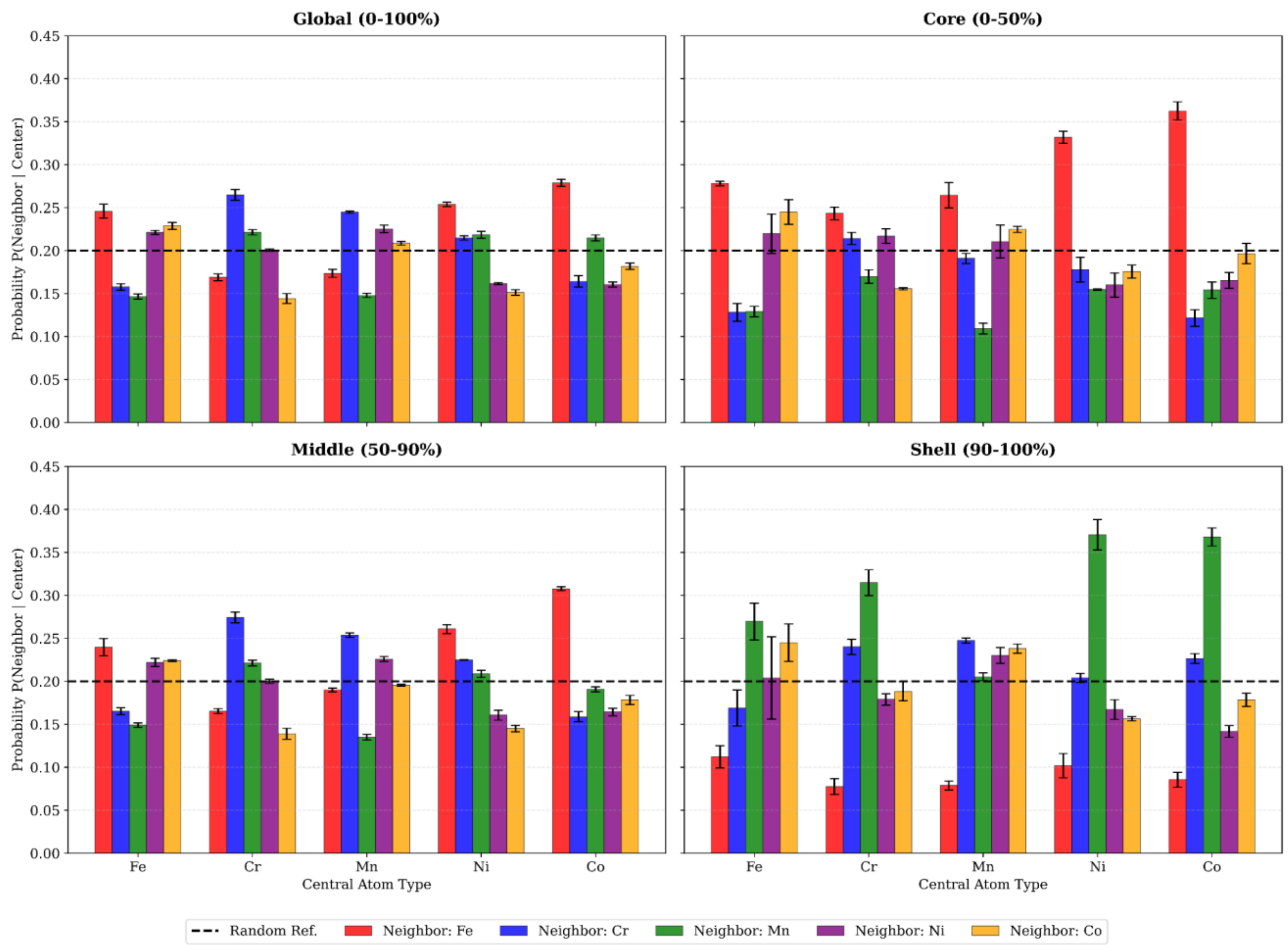

**Figure S5:** Conditional neighbour probability analysis characterizing the Chemical Short-Range Order (CSRO). The analysis highlights the formation of a cohesive Fe-rich core and a Mn-rich shell.

## 6. Thermodynamic calculations

In the present study, thermodynamic calculations were performed using the FactSage thermochemical software to investigate the phase evolution and stability of the CoCrFeMnNi high-entropy alloy (HEA). FactSage determines thermodynamic equilibrium by minimizing the total Gibbs free energy of the system. The calculations were carried out over a temperature range of 400–800 °C at a constant pressure of 1 atm. To evaluate the phase evolution within this temperature interval, all thermodynamically stable phases were considered, including possible two-phase and three-phase immiscibility regions. The calculations were performed under the assumption that the HEA system exists in thermodynamic equilibrium.